\documentclass[11pt,twoside]{article}

\usepackage{asp2006}
\usepackage{epsf}
\usepackage{psfig}
\usepackage{lscape}

\newcommand{\taud}{\tau_{\rm d}}
\begin{document}
\title{Interstellar Dust Models and Evolutionary Implications}%%% Fill in title
\author{B. T. Draine}   %%% Fill in author names
\affil{Princeton University}  %%% Fill in author affiliations

\begin{abstract} %%% Abstract to run on from here
The wavelength dependences of interstellar extinction and
polarization, supplemented by observed elemental abundances and 
the spectrum of infrared emission from
dust heated by starlight, strongly
constrain dust models.  One dust model that appears to be consistent
with observations is presented.  To reproduce the observed extinction,
the model consumes the bulk of
interstellar Mg, Si, and Fe (in amorphous silicates), 
and a substantial fraction of C (in carbonaceous material), with
size distributions and alignment adjusted to match observations.
The composition, structure, and size distribution of
interstellar grains is the result of injection of dust from stellar
outflows into the interstellar medium (ISM), followed by destruction,
growth, coagulation, and photoprocessing of interstellar grains.  The
balance among these poorly-understood processes is responsible for the
mix of solid material present in the ISM.  
Most interstellar grain material present in the diffuse ISM
must be grown {\it in} the ISM.
The amorphous silicate and
carbonaceous materials that form the bulk of interstellar dust 
must therefore be the
result of grain growth in the presence of ultraviolet radiation.
Dust in high-$z$ systems such as J1148+5251 is
also produced primarily
in the ISM, with supernova-produced dust contributing only
a small fraction of the total dust mass.
\end{abstract}

%%% MAIN BODY OF TEXT GOES HERE. CONSULT "INSTRUCTIONS FOR AUTHORS USING
%%% LATEX2E MARKUP", SECTIONS 2.3-2.6 FOR HELP WITH EQUATIONS, FIGURES,
%%% AND TABLES.

\section{Introduction}
\citet{Trumpler_1930} and \citet{Lindblad_1935}
recognized that reddening of distant stars was produced by submicron particles
present in  interstellar space,
but over 70 years later
we continue to work to determine the composition, shape, and
size distribution of interstellar dust.  Despite substantial advances in
our understanding of interstellar physics, the problem
is far too difficult for an {\it ab initio} approach.  We do not know enough
about physical conditions in stellar outflows (where some grains are formed)
and the interstellar medium (where grains undergo modification), and our
understanding of the physical and chemical processes occuring on grain
surfaces and within grains is very limited. 
We cannot predict with any
confidence how grain materials will evolve over hundreds of Myr in the 
interstellar medium (ISM).
We need to use what physics we know, but observations must guide us.

Almost all of the observational information concerning interstellar dust
arises from interaction of electromagnetic radiation with dust:
scattering, absorption, and emission.  The first ``detection'' of interstellar
dust was through extinction of starlight: \citet{Herschel_1785} remarked on
the absence of stars in certain portions of the Milky Way 
({\it ``An Opening in the Heavens''}), although he did not understand its
cause.  
Over a century later, 
\citet{Barnard_1907} noted that the
neighborhood of $\rho$~Oph was an example of 
``an apparently absorbing medium''.
We now study scattering, absorption, and emission of radiation by interstellar
dust at wavelengths ranging from \AA\ (X-rays) to cm (microwaves).

\section{Observational Constraints}
\subsection{Extinction}

\begin{figure}
\plotfiddle{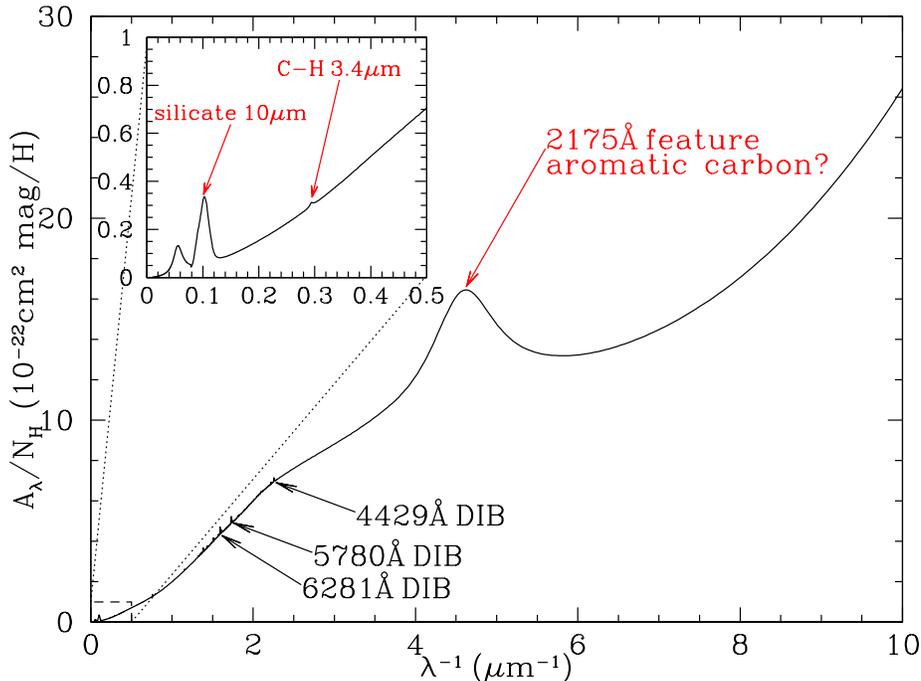}{8.0cm}{270}{45}{45}{-180}{255}
\caption{\label{fig:extcurv}Average extinction vs. $\lambda^{-1}$ for
dust in interstellar diffuse clouds.  For $\lambda^{-1} > 1\micron^{-1}$, this
is based on the parameterization of extinction
by \citet{Fitzpatrick_1999}, with addition of the 77 strongest 
diffuse interstellar bands (DIBs) from \citet{Jenniskens+Desert_1994}.
Three of the strongest DIBs are labelled.}
\end{figure}

The wavelength-dependent extinction of starlight -- the
so-called ``extinction curve'' -- remains
the principal source of information about interstellar dust.
Interstellar extinction has been studied on many sightlines at wavelengths
between about 2$\mu$m and 0.1$\mu$m.  On sightlines with sufficient
column density, it is possible to study the extinction at wavelengths as long
as $\sim20\mu$m. 
An average extinction curve is shown in Figure \ref{fig:extcurv}.

The extinction curve contains spectral features that constrain the composition
of the dust.  The
strongest feature by far is a broad ``bump'' peaking near 2175\AA.
The strength of this feature requires that it be produced by a substance
composed of high-abundance
elements, such as C, Mg, Si, or Fe \citep{Draine_1989a}.  
The position of the feature, and its width, are strongly suggestive of
$\pi\rightarrow\pi^*$ excitations in aromatic carbon, such as graphite or
polycyclic aromatic hydrocarbons.
Some authors \citep[e.g.,][]{Draine+Li_2007} 
think that the feature is produced by the large population of
polycyclic aromatic hydrocarbons that is required to explain a number
of infrared emission features.

The extinction curve in Fig.\ \ref{fig:extcurv} shows a conspicuous
feature near 10$\mu$m that can be confidently attributed to Si-O
stretching modes in amorphous 
silicates; a companion feature at 18$\mu$m is attributed
to O-Si-O bending modes.  The absence of fine structure in the
profile indicates that the silicate material is amorphous -- 
\citet{Kemper+Vriend+Tielens_2005} place an upper limit of 2\% on the 
crystalline fraction.

There is a weak absorption feature at 3.4$\mu$m that is attributed to the
C-H stretch in hydrocarbon material.   
The nature of the hydrocarbon material
is controversial:
from the profile shape and strength,
\citet{Pendleton+Allamandola_2002} estimate that the carbonaceous material
is $\sim$85\% aromatic and $\sim$15\% aliphatic, whereas
\citet{Dartois+MunozCaro+Deboffle+dHendecourt_2004} conclude that
{\it at most} 15\% of the carbon is aromatic.

Finally, we note that the extinction curve in Fig.\ \ref{fig:extcurv}
actually contains $\sim$200 weak but detected spectral features -- the
``diffuse interstellar bands'', or DIBs.
86 years after the first DIBs were reported by \citet{Heger_1922}, 
it is embarassing to have to admit that not a
single one has yet been identified.

The extinction curve shows a smooth rise from the near IR to the
vacuum UV.  A broad range of grain sizes is required to reproduce this: 
the rapid rise in extinction shortward of 1500\AA\
requires very large numbers of grains with radii $a < 0.02\,\mu$m, but the
extinction in the visible requires that most of the grain mass be
in grains with sizes $0.05\mu{\rm m}< a < 0.3\mu{\rm m}$.
The overall strength of the extinction per H nucleon requires that
the grains providing the bulk of the extinction be based on high-abundance
elements such as C, O, Mg, Si, and Fe.

Extinction curves vary from one sightline to another.
Figure \ref{fig:extcurv} shows an average extinction curve, but grain models
must be able to account for the 
observed sightline-to-sightline variations in the 
wavelength-dependence
of extinction.

\subsection{Other Evidence}

In addition to the wavelength-dependent extinction, a modern dust model must
also be consistent with a wide range of other observations of electromagnetic
radiation absorbed, scattered, or radiated by dust:\\
\smallskip
\hspace*{1em}$\bullet$ Polarization of starlight by interstellar grains.\\
\smallskip
\hspace*{1em}$\bullet$ Scattering of starlight by grains (reflection nebulae, 
diffuse galactic light).\\ \smallskip
\hspace*{1em}$\bullet$ Small-angle scattering of X-rays by interstellar 
grains.\\
\smallskip
\hspace*{1em}$\bullet$ Infrared and submm emission from interstellar grains 
heated by starlight.\\
\smallskip
\hspace*{1em}$\bullet$ Polarization of the infrared and submm emission from 
dust.\\
\smallskip
\hspace*{1em}$\bullet$ Microwave emission from dust.\\
In addition to the above, we can study the interstellar grains that are
entering the heliosphere today
\citep{Landgraf+Baggaley+Grun+etal_2000,
       Kruger+Landgraf+Altobelli+Grun_2007}, 
and we can analyze presolar grains trapped 4.5 Gyr ago
in primitive meteorites and comets.

Our estimates
of the total interstellar abundances of elements such as C, Mg, Si, and Fe,
plus direct measurement of gas-phase abundances using
absorption lines, tell us how much of each element appears to be
locked up in solid grains.
Gas-phase abundances of species such as Si, Ca, and Ti are observed to
vary from one sightline to another, which implies that depletion from the
gas must
be able to occur relatively rapidly; a successful grain model must
provide enough surface area for
depletion of Si$^+$, Ca$^+$, and Ti$^+$ to take place rapidly.

The observed sightline-to-sightline variations of gas-phase 
D/H is best understood
if D is actually depleting onto dust grains
\citep{Draine_2004a,Draine_2006b} and evidence for this is now
strong
\citep{Prochaska+Tripp+Howk_2005,
       Linsky+Draine+Moos+etal_2006,
       Lallement+Hebrard+Welsh_2008}.
A dust model must contain a component capable of trapping a substantial
fraction of the D present in the ISM.
The PAHs may be able to accomplish this \citep{Draine_2006b}.

In general, 
dust grains play a crucial role in interstellar chemistry, and a successful
dust model should eventually be able to ``predict'' the rate for formation
of interstellar H$_2$ via reactions on grain surfaces.  Photoelectrons
emitted by interstellar grains make a major contribution to heating of
interstellar gas, and a successful grain model must be in quantitative
agreement with heating rates inferred from observations of interstellar gas.

\section{A Model Using Amorphous Silicate and Carbonaceous Grains}

Current modeling techniques do not allow us to
``invert'' the observational constraints to arrive at a grain model.
Constrained by elemental abundances, 
all sensible grain models consist predominantly of 
carbonaceous and amorphous silicate material,
but there are various candidates for the carbonaceous material.
For example, \citet{Zubko+Dwek+Arendt_2004}
consider models involving varying amounts of
PAHs, graphite, 3 different types (AC, BE, ACH2) of amorphous carbon,
and an organic refractory material with composition C$_{25}$H$_{25}$O$_5$N.

Here we will focus on one particular model, in which the carbonaceous
material in ultrasmall particles is assumed to have the optical and thermal
properties
of PAHs, while the larger carbonaceous grains are assigned the optical
properties of graphite.
If polarization is not of interest, the particles may be
approximated as spheres.
For these materials,
\citet[][hereafter WD01]{Weingartner+Draine_2001a} 
found size distributions that reproduce
the extinction in various regions of the Milky Way, and in the Large and
Small Magellanic Clouds.
For current estimates of solar and interstellar abundances, the WD01 size
distributions
use somewhat more C, Mg, and Si than is nominally available, but one should
keep in mind that (a) there are modeling uncertainties (e.g., the grains are 
certainly not solid spheres), (b) the actual interstellar abundances
of C, Mg, Si, and Fe are uncertain,
(c) gas-phase abundances of C have recently been revised downward
\citep{Sofia+Parvathi_2009}, increasing the amount of carbon inferred to
be in dust: it now appears that only $\sim1/3$ of the C is in the gas,
and 2/3 is in the dust.

%%%%%%%%%%%%%%%%%%%%%%%%%%%% figure 2 %%%%%%%%%%%%%%%%%%%%%%%%%%%%%%%%%%%%%%%%
\begin{figure}
% draine_f2.eps = /u/draine/work/papers/sedfit/n7331.eps
\plotfiddle{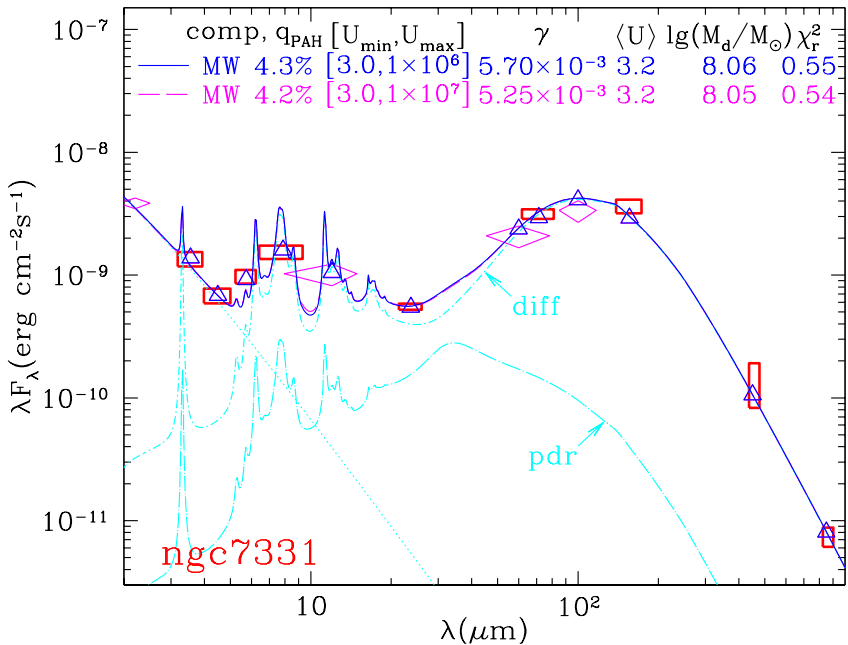}
{8.0cm}{270}{42}{42}{-190}{240}
\\
\caption{\label{fig:ngc7331}Global emission from NGC 7331
\citep[from][]{Draine+Dale+Bendo+etal_2007}.  
Symbols are 2MASS, Spitzer, IRAS, and SCUBA photometry.
The solid curves is generated by the DL07 dust model, which
requires a dust mass $1.1\times10^8M_{\sun}$ to reproduce
the observed emission spectrum.}
\end{figure}
%%%%%%%%%%%%%%%%%%%%%%%%% end figure 2 %%%%%%%%%%%%%%%%%%%%%%%%%%%%%%%%%%%%%%%%
%%%%%%%%%%%%%%%%%%%%%%%%%%%% figure 3 %%%%%%%%%%%%%%%%%%%%%%%%%%%%%%%%%%%%%%%%%
\begin{figure}
%/u/draine/work/papers/dnf08/nucspec_fig1a.eps = draine_f3.eps
\plotfiddle{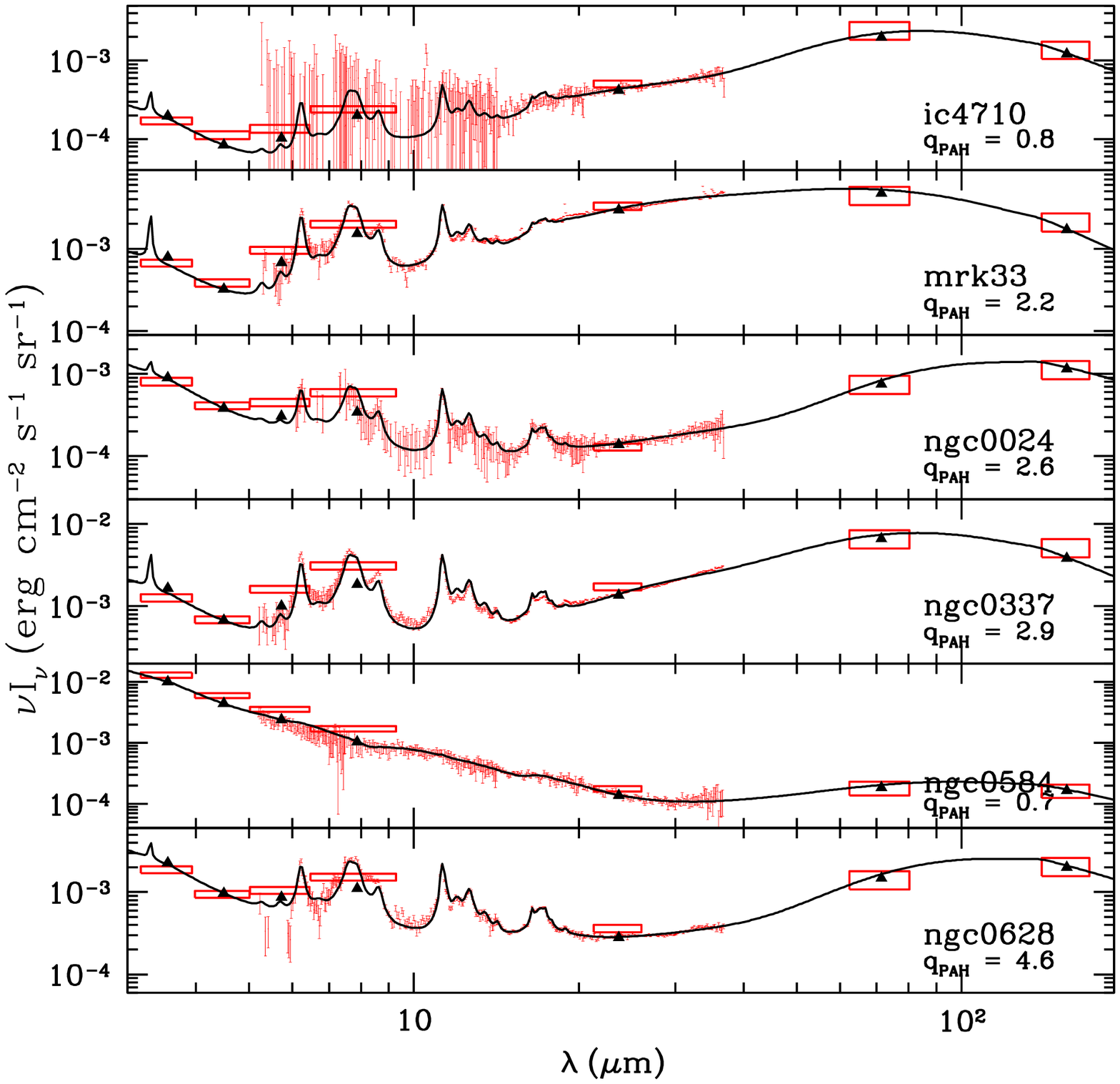}
{7.9cm}{0}{69}{69}{-200}{-140}
\vspace*{1.5em}
\caption{\label{fig:irspec}Emission from central regions of 6 SINGS galaxies.
Error bars are for 5--33$\mu$m IRS spectra
\citep{Smith+Draine+Dale+etal_2007}.
Rectangles are IRAC and MIPS photometry.
Solid line is a model consisting of starlight plus emission from dust heated
by starlight.  The DL07 physical grain 
model is able to reproduce both the PAH features and
the IR continuum for diverse galaxies, but with different PAH fractions 
$q_{\rm PAH}$ and different starlight intensity distributions.
From \citet{Draine+Reyes+Smith+etal_2009}.}
\end{figure}
%%%%%%%%%%%%%%%%%%%%%%%%%% end figure 3 %%%%%%%%%%%%%%%%%%%%%%%%%%%%%%%%%%%%%%%

The PAH-graphite-amorphous silicate model satisfactorily
reproduces interstellar extinction (see WD01), 
but it can also be tested in other ways.
A dust grain illuminated by starlight will absorb energy, which will
be reradiated in the infrared.  The infrared
emission spectrum will depend upon both the temperature of the grain and
its infrared opacity.  

Large $a\ga 0.02\micron$ grains
in the local starlight background will be heated to a more-or-less
steady temperature of 15--20K.  However, very small grains (1) absorb photons
much less frequently, and (2) have very small heat capacities, so that one
absorbed photon can raise the grain to a high temperature, followed by
very rapid cooling.  
Because of this, one cannot speak of an average temperature
for very small grains -- instead, one must determine the
temperature distribution
function $(dP/dT)$, where $dP$ is the probability of finding the grain in
the temperature interval $[T,T+dT]$.  
The infrared emission is given by
\begin{equation}
\frac{I_\nu}{N_{\rm H}} = \sum_{{\rm comp}\,i}~\sum_{{\rm size}\,j} 
\frac{n_{ij}}{n_{\rm H}}
\int dT \left(\frac{dP}{dT}\right)_{ij} B_\lambda(T) C_{{\rm abs},ij}(\lambda)
~~~.
\end{equation}
To calculate the emission from a grain
model that includes very small grains, one must first find $(dP/dT)_{ij}$ for
the different compositions $i$ 
and grain sizes $j$ that are present in the model.
For large grains, $dP/dT \rightarrow \delta(T-\langle T\rangle)$, 
where $\delta$ is
the Dirac delta function, but for small grains $dP/dT$ has a tail extending
to high $T$.

The DL07 grain model uses the WD01 size distribution for all except the
PAHs; the size distribution of the PAHs was adjusted to try to reproduce
the $\lambda < 25\micron$ infrared emission.
DL07 have solved for the 
temperature distribution functions $dP/dT$ for different grain sizes and
composition,
and have calculated the resulting infrared emission spectrum for
various intensities of the starlight heating the grains.

The diffuse ISM away from the Galactic plane in the
Milky Way has been measured by COBE-DIRBE
\citep{Dwek+Arendt+Fixsen+etal_1997,
       Arendt+Odegard+Weiland+etal_1998} 
and by COBE-FIRAS
\citep{Finkbeiner+Davis+Schlegel_1999}.
The emission predicted for the DL07 grain model is in good agreement with
the observed diffuse emission from high-latitude regions in the Milky Way.

If we assume that grains in other galaxies
are like grains in the Milky Way, we can test the grain model by comparing
it to the emission observed from other galaxies.
The Spitzer Infrared Nearby Galaxy Survey
\citep[SINGS,][]{Kennicutt+Armus+Bendo+etal_2003} 
consists of 75 nearby galaxies observed
with all of the instruments on Spitzer Space Telescope
\citep{Werner+Roellig+Low+etal_2004}.
\citet{Draine+Dale+Bendo+etal_2007} have used the DL07 dust
models to estimate the dust content and starlight intensities for 65 of the
SINGS galaxies.
Figure \ref{fig:ngc7331} shows the observed SED for one example, 
the well-observed Sb galaxy NGC~7331.
With suitable assumptions for the starlight heating the dust, the DL07 dust
model closely reproduces the observed infrared and submm emission.

Modeling the IR emission from NGC~7331 yields an estimated dust mass
$M_{\rm dust}=1.1\times10^8M_{\sun}$.
The H\,I\,21cm emission give an HI mass $M({\rm HI})=1.0\times{10}^8M_{\sun}$,
and the CO~1$\rightarrow$0 luminosity gives an estimated H$_2$ mass
$M({\rm H}_2)=1.6\times10^{10}M_{\sun}$ (for an assumed CO ``X factor''
$4\times10^{22}{\rm cm}^{-2}\,{\rm K~km~s}^{-1}$), resulting in a dust-to-H 
mass ratio $M_{\rm dust}/M_{\rm H}=0.0043$, reasonably
close to the expected value $\sim 0.007$, especially when one considers
the uncertainties in estimation of the H$_2$ mass from the CO~1$\rightarrow$0
luminosity.

\section{Polarization of Starlight and Polarized IR-Submm Emission}

WD01 showed that a model based on amorphous silicate
and carbonaceous grains, including PAHs, could reproduce observations of
extinction in the Milky Way, LMC, and SMC, and DL07 showed that these same
dust models could reproduce observations of infrared emission.

In addition to interstellar extinction, a dust model should be able to
reproduce observations of polarization of starlight by aligned dust grains.
The infrared emission from aligned dust grains will be polarized, and
microwave and 
submillimeter polarimetry has advanced to the point where
it is beginning to be possible to measure the polarized emission from the
diffuse ISM.
For computational convenience, the WD01 and DL07 models assumed spherical 
grains, and therefore couldn't say anything about polarization.

Early modeling of starlight polarization at optical wavelengths
approximated the grains by
infinite cylinders \citep{Greenberg+Hong_1974}, but improved numerical
methods and faster computers allowed \citet{Kim+Martin_1995} to model the
grains as spheroids.
Advances in polarimetric techniques at submm to cm wavelengths now allow
detection of polarized emission from the diffuse ISM
\citep[e.g.][]{Page+Hinshaw+Komatsu+etal_2007} and theorists should now
make predictions for the degree of polarization that is expected for different
dust models.

In addition to reproducing the wavelength-dependent extinction,
a dust model must first reproduce the polarization of
starlight as a function of wavelength. 
All dust models must have aligned, nonspherical grains containing
amorphous silicates, because the interstellar silicate feature is known to
produce polarization in extinction
\citep{Smith+Wright+Aitken+etal_2000}. 
If carbonaceous material is present in
a separate grain population, at this time there is no evidence that such
grains are aligned.  Absence of polarization in the 3.4$\mu$m C-H stretch
feature suggests that the carbonaceous
particles (at least those responsible for the 3.4$\micron$ feature)
may {\it not} be aligned \citep{Chiar+Adamson+Kerr+etal_2005}.  

\begin{figure}
%/u/draine/work/papers/firpol/F7.eps = draine_f4a.eps
%/u/draine/work/papers/firpol/F8b.eps = draine_f4b.eps
\plotfiddle{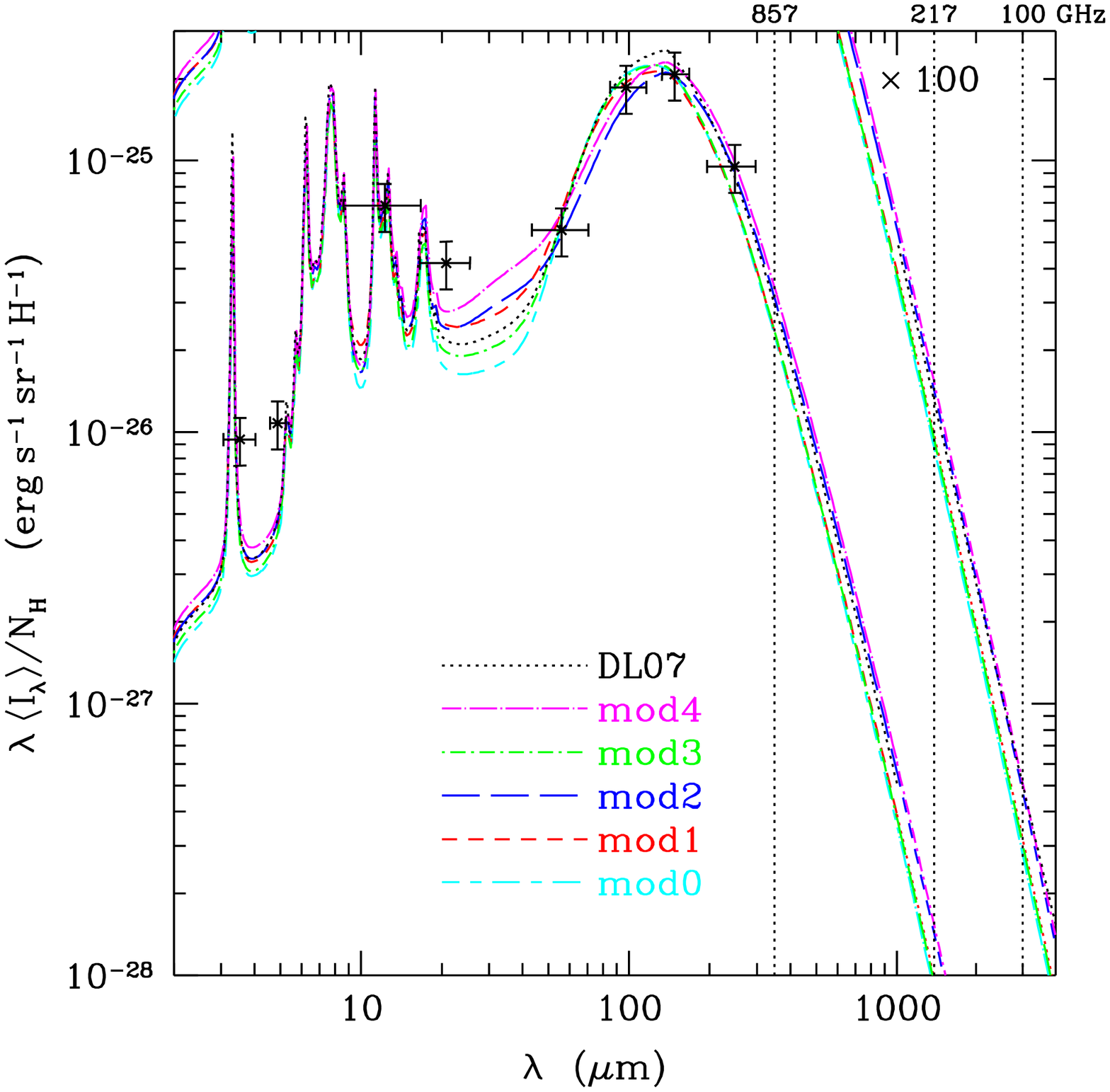}{8.0cm}{0}{35}{35}{-200}{-10}
\plotfiddle{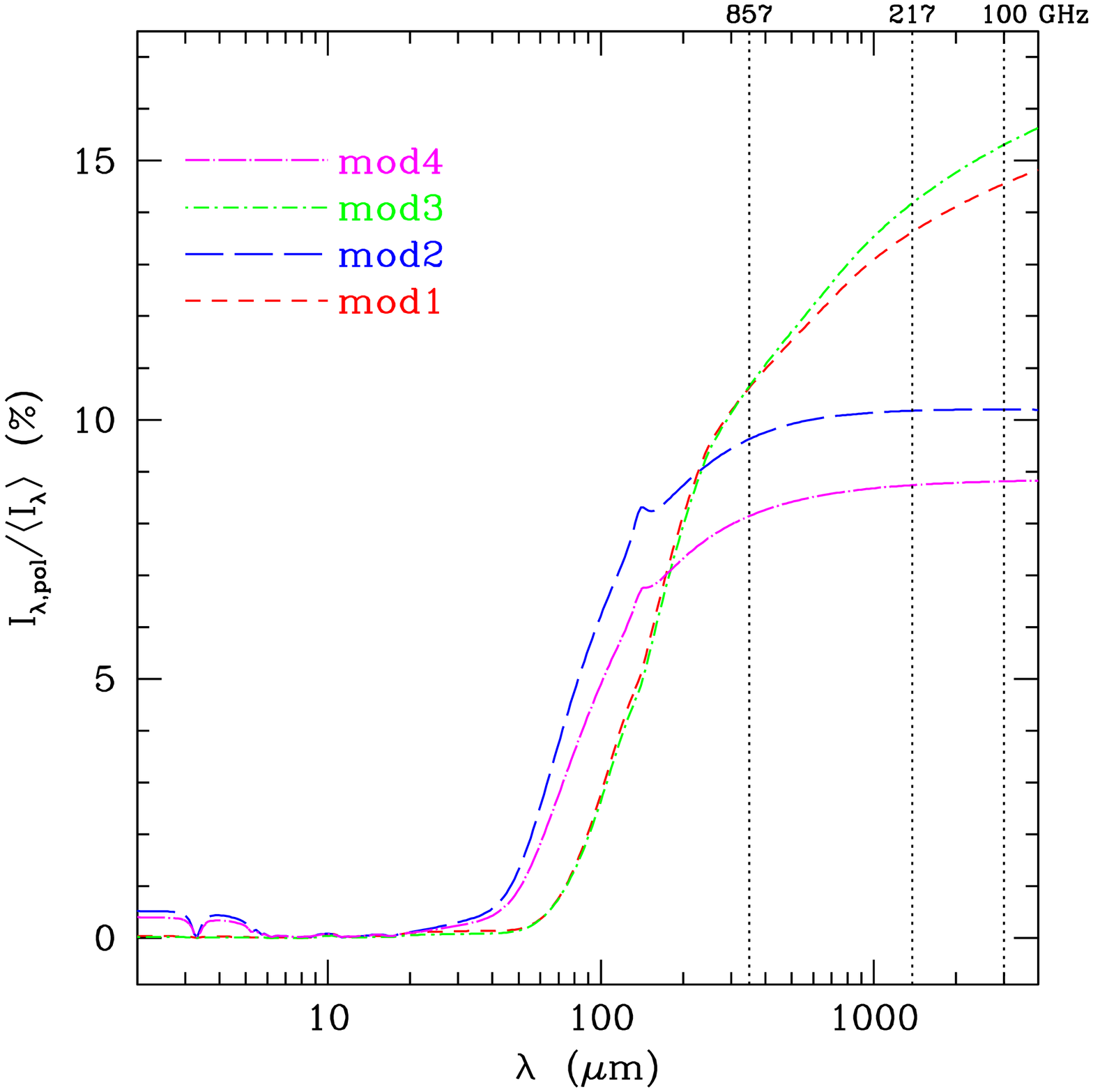}{8.0cm}{0}{35}{35}{-10}{230}
\vspace*{-10.0cm}
\caption{\label{fig:FIRpol}{\it Left:} Infrared emission 
vs.\ wavelength $\lambda$ for 4 models (mod1--mod4) with
spheroidal grains that reproduce
extinction and polarization of starlight \citep{Draine+Fraisse_2009}
and for
the spherical grain model of \citet[][DL07]{Draine+Li_2007}.
Crosses show the emission per H for local high-latitude
dust observed by DIRBE 
\citep{Dwek+Arendt+Fixsen+etal_1997,
       Arendt+Odegard+Weiland+etal_1998}.
{\it Right:} Linear polarization vs.\ $\lambda$ 
for emission perpendicular to the local magnetic
field. }
\end{figure}

Using models based on carbonaceous grains
(including PAHs) and amorphous silicate grains with various assumptions
about (1) the shapes of the grains and (2) whether or not the carbonaceous
grains are aligned,
\citet{Draine+Fraisse_2009} are able to find size distributions and
alignment functions (degree of alignment as a function of grain size) that
are consistent with observations of interstellar extinction and starlight
polarization.
Figure \ref{fig:FIRpol}({\it left}) 
shows the infrared emission from their models when
the dust is heated by the local interstellar radiation field.
The models are all in good agreement with the far-infrared emission per
H observed by DIRBE 
\citep{Dwek+Arendt+Fixsen+etal_1997,
       Arendt+Odegard+Weiland+etal_1998}.

Fig.\ \ref{fig:FIRpol}({\it right}) 
shows that the different models predict quite
different degrees of polarization, and with different 
dependences on wavelength.
This is good news -- it means that upcoming multiwavelength measurements of
linear polarization by {\it Planck} 
should be able to reject some (or perhaps all!) of these models.

\section{Where Do Interstellar Grains Come From?}

The evolution of interstellar dust involves a complex interplay of many 
poorly-understood processes, in an arena -- the ISM -- that we
also have a very incomplete picture of.
{\it Ab initio} approaches to the evolution of interstellar dust are therefore
premature.

Under these circumstances, the best approach to understanding
the evolution of interstellar dust is to first determine what kind of dust
is present in the ISM {\it today}.  We can then try to
construct an evolutionary scenario that is consistent with the grains
that are observed to be present today.

To summarize: 
Observations of interstellar gas-phase abundances (``depletions''),
extinction, polarization, and emission by interstellar dust, 
and models that are able to reproduce
those observations, tell us that:
\begin{enumerate}
\item The abundant elements Mg, Si, Fe reside primarily in dust.
\item Perhaps 2/3 of C is in dust.
\item The Si is predominantly in amorphous silicate material.  Based on
available abundances, the composition is likely to be (approximately)
Mg$_{1.1}$Fe$_{0.9}$SiO$_4$.
\item There is a substantial population of PAHs that contains
$\sim$10--20\% of the interstellar C in the Milky Way; 
$\sim$4--5\% of the total grain mass is
contributed by PAHs in the Milky Way and other star-forming galaxies
with $\sim$solar metallicity \citep{Draine+Dale+Bendo+etal_2007}.
\end{enumerate}
Whatever the processes of interstellar grain evolution, they must be able
to account for the above facts.
\subsection{Mass Budget for the Milky Way ISM}
\begin{figure}
% /u/draine/work/papers/dnf08/fmass.eps = draine_f5.eps
\plotfiddle{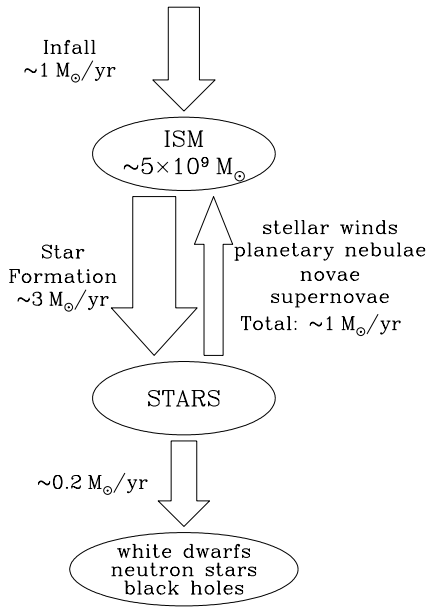}{8.0cm}{0}{30}{30}{-100}{0}
\caption{\label{fig:massflow}
         Baryon budget for the Milky Way disk (see also Table 1).}
\end{figure}

Figure \ref{fig:massflow} illustrates the 
overall flow of mass into and out of the ISM.
%%%%%%%%%%%%%%%%%%%%%%%%%%%%%%%% figure 5 %%%%%%%%%%%%%%%%%%%%%%%%%%%%%%%%%%%
%%%%%%%%%%%%%%%%%%%%%%%%%%%%%%%% end figure 5 %%%%%%%%%%%%%%%%%%%%%%%%%%%%%%
The mass of interstellar gas in the Milky Way is $\sim5\times10^9M_{\sun}$,
divided approximately equally between atomic and molecular gas.
Gas is being converted into stars at a rate of $\sim3M_{\sun}/{\rm yr}$.  
If there were no source of gas, 
the ISM would be consumed on a relatively short timescale 
$\sim5\times10^9M_{\sun}/(3\,M_{\sun} {\rm yr}^{-1}) 
\approx 1.7\times10^9$~yr.
However,  there is
return of material to the ISM from stars via various kinds of stellar outflows
(stellar winds, planetary nebulae, novae, supernovae) and, in addition,
there is evidence for infall of low metallicity gas.
The rates are uncertain, particularly for infall, but it appears (see
Figure \ref{fig:massflow}) that
the {\it net} rate at which the ISM is losing mass 
is $\sim1 M_{\sun}$/yr, so that
the characteristic decay time is of order $5\times10^9$~yr, 
which seems a plausible
number for a galaxy that has been making stars in the disk at a more-or-less
steady rate over the last $\sim$9 Gyr 
\citep{Rocha-Pinto+Scalo+Maciel+Flynn_2000}.

\section{Production of Stardust}

Some of the stellar outflows returning matter to the ISM are dusty.
Outflows from cool oxygen-rich red giants contain silicate grains,
and the outflows from cool carbon stars contain solid carbonaceous grains,
in some cases with SiC as one of the carbon-bearing materials.
The ejecta that form planetary nebulae are generally dusty.
Some novae produce dust, but the net contribution from novae 
appears to be small.

At least some stardust comes from supernovae: 
supernova-produced dust grains have
been identified in meteorites, and the internal extinction and IR
emission from some young supernova remnants is evidence for dust formation.
Supernovae have even been described as ``dust factories''
\citep{Sugerman+Ercolano+Barlow+etal_2006}.
However, the average mass of dust per SN 
does not appear to be large: the Type II
SN 1987a produced $\la 8\times10^{-4}M_{\sun}$ of dust
\citep{Ercolano+Barlow+Sugerman_2007}.
The current record-holder for dust production is the Type II SN 2003gd, which 
produced $\la 0.02 M_{\sun}$ of dust 
\citep{Sugerman+Ercolano+Barlow+etal_2006}.
There is as yet no evidence of dust formation in Type Ia SNe.
While only a very small number of SNe have as yet been studied, it appears
that SNe produce, on average, 
$\la 0.01M_{\sun}$ of dust, containing $\la10\%$ of the
condensible elements.
It should also be kept in mind that the dust that has been detected in these
supernovae is moving at high speeds ($\sim 10^3\,{\rm km~s}^{-1}$), 
and is expected to undergo erosion by sputtering and
possible destruction in 
grain-grain collisions when it is eventually decelerated to the low velocities
of the quiescent ISM.
While dust production unquestionably occurs, 
``dust factory'' may be a misnomer.

\begin{table}[!ht]
\caption{Injection of Gas and Stardust from Stellar Sources}
\smallskip
\begin{center}
{\small
\begin{tabular}{ccl}
\tableline
\noalign{\smallskip}
gas & dust & Stellar Source\\
($M_{\sun}/$yr) & ($M_{\sun}/$yr) &\\
\noalign{\smallskip} 
\tableline
\noalign{\smallskip}
0.4 & 0.002      & Planetary Nebulae ($\sim$0.3/yr)\\
0.5 & 0.0025      & Red Giant, Red Supergiant, C star winds\\
0.06 & $<$0.0001? & OB, WR, other warm/hot star winds\\
0.25 & \,\,\,\,0.0002?    & SNe (1/100~yr, $\sim10^{-2}M_{\sun}$ dust/SN?)\\
0.01 & \,\,0.00001    & Novae (100/yr, $10^{-7}M_{\sun}$ dust/nova?)\\
\tableline
{\bf$\sim$1.2} & {\bf$\sim$0.005}~~ & {\bf All stellar sources}\\
\tableline
\noalign{\smallskip}
\end{tabular}\\
}
\end{center}
\end{table}

An accurate inventory of the different sources of stardust is not available,
but provisional estimates are offered in Table 1.
The overall
gas return rates from the different stellar sources in Table 1
can be estimated reasonably reliably, given our knowledge of the IMF
\citep{Kroupa_2001}:
(1) stars with initial masses above $\sim8M_{\sun}$ become supernovae,
leaving behind either neutron stars or black holes.
(2) Initial masses between $\sim 0.9 M_{\sun}$ and $\sim 8 M_{\sun}$ leave
behind white dwarfs of varying masses.
(3) Stars with initial masses below $\sim 0.9M_{\sun}$ do not evolve on the
timescales of interest.

We do not have a good understanding of either the stellar wind phenomenon
itself, or the formation of dust in some stellar winds, so the estimates
of dust injection are less reliable.  The estimates for planetary
nebulae and winds from cool stars assume that the dust mass is 0.5\% of the
gas mass, which corresponds to efficient condensation of the condensible
elements.
According to the estimates in Table 1, stardust is entering the ISM at a rate
$\dot{M}_{\rm stardust}\approx 0.005 M_{\sun}$/yr.

\section{Stardust Lifetimes}

Newly-made stardust moves away from its stellar source and
begins a journey through the ISM.
If nothing else happens, the grain will eventually find itself in a parcel
of gas collapsing to form a new star.
If this were the only process that removed grains from the ISM, 
the grain lifetime $\taud$ 
would be the lifetime of material against star formation:
$\taud = \tau_{\rm SF}\approx M_{\rm ISM}/{\rm SFR} \approx 2\times10^9$~yr.

The ISM is a violent neighborhood, agitated by fast
stellar winds and supernova explosions.  
A grain that is overrun by a fast shock is subject to sputtering by the
shocked gas.
Sputtering in shock waves has been studied by a number of authors.
\citet{Draine+Salpeter_1979b} found that $a=0.1\micron$ silicate grains
had $>50\%$ of the grain material returned to the gas in radiative
shocks with shock speeds $v_s > 200\,{\rm km\,s}^{-1}$.

We can estimate the global effects of SNe by a very simple argument.
A supernova blastwave with initial energy $E_0=10^{51}$erg in gas with
density $n_{\rm H}\approx 1\,{\rm cm}^{-3}$ is Sedov-like until the
shock speed drops to $\sim 200\,{\rm km\,s}^{-1}$.  During the Sedov
phase,
\begin{equation}
M v_s^2 \approx E_0 ~~~,
\end{equation}
and therefore the mass of gas shocked at shock speeds $> 200\,{\rm km s}^{-1}$
is
\begin{equation}
M(v_s > 200{\rm km s}^{-1}) \approx 
\frac{10^{51}{\rm erg}}{(200\,{\rm km s}^{-1})^2} \approx 1260 M_{\sun}
~~~.
\end{equation}
Thus one supernova can ``clean'' $\sim1000\,M_{\sun}$ of ISM of dust.
If $E_0=10^{51}$erg is a representative value for the kinetic energy of
a SN, and the SN rate in the Milky Way disk is $\sim1/(100\,{\rm yr})$, then
the probability per unit time of an interstellar grain being overtaken
by a $v_s > 200\,{\rm km\,s}^{-1}$ shock is
\begin{equation}
\taud^{-1} \approx \frac{1260 M_{\sun}/100\,{\rm yr}}{5\times10^9M_{\sun}}
\approx \frac{1}{4\times10^8 {\rm yr}}
~~~.
\end{equation}

There are other destructive processes as well, particularly grain-grain
collisions in slower shock waves.
Detailed studies 
\citep{Barlow+Silk_1977,
       Draine+Salpeter_1979b,
       Dwek+Scalo_1980,
       Jones+Tielens+Hollenbach+McKee_1994}
that try to take into account
the complications of a multiphase
ISM find solid matter lifetimes $\taud \approx 4\times10^8$yr,
consistent with the simple estimate above.

To restate this: for a randomly-selected Si atom in a grain,
the probability per unit time that it will be returned to the gas is
$\taud^{-1}\approx 1/(4\times10^8{\rm \,yr})$.
In steady state, stardust injection = stardust removal, and 
we can estimate the mass of stardust that should be present in the ISM.
\begin{equation}
\dot{M}_{\rm stardust} \approx M_{\rm stardust}
\left[\tau_{\rm SF}^{-1} + \taud^{-1}\right] ~~~,
\end{equation}
from which we estimate $M_{\rm stardust} \approx 1.6\times10^6 M_{\sun}$.
However, the total mass of refractory interstellar dust is
$M_{\rm ismdust}\approx 0.005\times M_{\rm ISM}
\approx 2.5\times10^7 M_{\rm sun}$.
Thus we conclude that 
\begin{equation}
\frac{M_{\rm stardust}}{M_{\rm ism dust}} \approx 
\frac{1.6\times10^6M_{\sun}}{2.5\times10^7M_{\sun}}
\approx 0.06 ~~~:~~~
\end{equation}
\\
\centerline{\fbox{\it Most interstellar dust is not stardust.}}\\
\\
Stardust accounts for only $\sim4\%$ of the total mass
of interstellar dust.

Another way to approach the problem is to estimate the gas-phase abundance
of Si. Let $f_{\star 0}({\rm Si})$ be the fraction of Si entering the ISM
from stellar sources (stellar winds, planetary nebulae, supernovae...) 
 that is in solid form.
Because supernovae appear to condense $\la 10\%$ of their condensibles
into grains \citep{Ercolano+Barlow+Sugerman_2007} and
the winds from many stars (e.g., O stars) appear to be dustless,
we may estimate $f_{\star_0}({\rm Si}) < 0.5$.
The fraction of interstellar Si that is in stardust will be
\begin{equation} \label{eq:f(Si)}
f_\star({\rm Si}) = \frac{f_{\star 0}({\rm Si})}{1 + t_{\rm SF}/t_{\rm d}}
\approx 0.2 f_{\star 0}({\rm Si}) \approx 0.1 ~~~.
\end{equation}
But in fact interstellar Si is almost always found to be strongly
depleted, with $f({\rm Si}) \approx 0.9$ in dust.
Therefore, we reach the same conclusion as before: most of the Si in 
interstellar grains is {\it not} in stardust -- 
it must have been added to the grains {\it in the ISM}.

This is not a new result -- \citet{Draine+Salpeter_1979b} reached the
conclusion that grain destruction was rapid and that regrowth of dust in
the ISM was required to explain the observed depletions.
The numbers basically haven't changed appreciably since then; 
the argument has been reiterated
a number of times
\citep[e.g.,][]
      {Dwek+Scalo_1980,
       Draine_1990,
       Weingartner+Draine_1999,
       Draine_2006c}.
Nevertheless, some authors 
continued to hold the view that the solids in the ISM
were primarily formed in stars.

For example, a number of papers have argued that the low abundance of
crystalline silicate material in the ISM requires that
crystalline stardust material must be amorphized by cosmic ray damage
in the ISM
\citep[e.g.][]{Bringa+Kucheyev+Loeffler+etal_2007}.  However, the low
abundance of crystalline silicates in the ISM can be easily understood:
(1) grain destruction reduces the abundance of
stardust in the ISM, and (2) the silicate material grown in the
ISM is amorphous.

Isotopically anomalous silicates found in interplanetary dust particles (IDPs)
are presumably stardust.
Techniques have recently been developed for study of stardust silicates
in meteorites 
\citep[e.g.][]{Nguyen+Zinner_2004b,
               Keller+Messenger_2008}.
Recent studies of stardust silicates identified in this way find that
it is $\sim$80\% amorphous, and $\sim$20\% crystalline material
(L.~P.~Keller 2008, talk given at this
conference).
If this 4:1 amorphous:crystalline ratio applies to the overall production of
silicate stardust, then the observed
upper limit of $\sim$2\% on the crystalline fraction in the ISM
\citep{Kemper+Vriend+Tielens_2005}
implies an upper limit of $2\%\times(4+1)=10\%$ on the fraction of 
interstellar
silicate material contributed by silicate stardust.
This is in agreement with our estimate in eq.\ (\ref{eq:f(Si)}).
The low crystalline fraction in interstellar silicates does not require
that some of the crystalline material be amorphized -- 
the expected destruction
processes in the ISM suffice to keep the crystalline fraction
below the current upper limit.

\section{Growth of Solid Material in the ISM}
If most interstellar solid matter is not stardust, it follows that\\
\\
\centerline{
\fbox{\it Most of the material in interstellar grains was formed in the ISM.}}
\\
\\
The growth of grain material takes place on the surfaces of already-present
grains -- a process also referred to as ``depletion'' of condensible
elements from the gas.  We know that this process takes place, because
we observe sightline-to-sightline variations in the gas phase abundances of 
the ``depletable'' species.\footnote{
  Mg, Al, Si, P, Cl, Ca, Ti, Cr, Fe, and Ni are the prime examples of elements
  whose gas-phase abundances show order-of-magnitude variations from one
  sightline to another, the result of depletion onto grains.}

The depletion process has three stages:
\begin{enumerate}
\item An atom or ion of the element
must collide with a preexisting grain surface.
\item The atom must
become bound to the grain material in a way that allows it to be retained
in the presence of fluxes of ultraviolet radiation and reactive species
such as atomic H and O.
\item The resulting grain material(s) will undergo heavy UV irradiation -- 
whatever survives (or is produced by) UV photolysis will be what is 
present in the ISM today.  
Cosmic ray irradiation also occurs, but is of secondary
importance.
\end{enumerate}
Step 1 is just a matter of kinetics: 
The rate for a given ion to hit a grain 
will be determined by the total
geometric area provided by the grains -- the surface area will be dominated
by the smallest grains -- and by
the random velocities of atoms and ions and
the (possibly large) velocities of grains relative to the gas\footnote{
  \citet{Yan+Lazarian+Draine_2004} conclude that MHD turbulence will
  cause $a> 0.04\mu$m silicate grains
  in the ``cold neutral medium'' 
  ($n_{\rm H}\approx 30\,{\rm cm}^{-3}$, $T\approx 100$\,K) 
  to move relative to the gas
  with velocities $\ga 1$\,km/s, large compared to
  the mean thermal speed $(8kT/\pi m)^{1/2} = 0.27$\,km/s of a Si ion.}.
In diffuse regions (where UV radiation is
present) the ion-grain collision rate
will be affected strongly by the charge state of the grains.
Collisions
between positively charged grains and ions such as Si$^+$ will be 
suppressed, but part of the grain population will be neutral, and the
smaller PAHs are expected to have an appreciable fraction that are negative,
with which ions will have enhanced collision rates.

Let $\Sigma_{{\rm d},21} \cdot 10^{-21}{\rm cm}^2/{\rm H} =$ dust geometric
cross section per H nucleon.  From modeling the UV extinction, we know that
$\Sigma_{{\rm d},21}\ga 1$.
Let $v_a$ be the velocity of an atom relative to the grains.
The timescale $\tau_{\rm acc}$ for this atom to collide with a grain
surface is
\begin{equation}
\tau_{\rm acc} = \frac{1}{n_{\rm H}\Sigma_d v_a}
= 1.0\times10^7{\rm yr}\left(\frac{30{\rm cm}^{-3}}{n_{\rm H}}\right)
\frac{1}{\Sigma_{\rm d,21}}
\left(\frac{{\rm km~s}^{-1}}{v_a}\right) ~~~.
\end{equation}
This time is relatively short. In the case of ions, the accretion rate
will be modified by Coulomb repulsion (by positive grains) or attraction
(by negative grains).  
\citet{Weingartner+Draine_1999} discussed 
the kinetics of accretion of ions (Ti$^+$ was used as an example) onto
grains in diffuse interstellar clouds, with attention to the charge
distribution of the PAHs.  They found that the very large surface
area of PAHs resulted in a depletion rate
\begin{equation}
\tau_{\rm acc} \approx 2 \times 10^5 {\rm yr} ~~~.
\end{equation}
The fact that this is a factor 50 smaller than eq. (8) can be attributed to
two factors.  First, the total surface area in small particles is 
considerably  
larger than $10^{-21}{\rm cm}^2/$H.
Second,
carbonaceous grains smaller than $10^{-6}\,$cm
and silicate grains smaller than $2\times10^{-6}\,$cm were negatively
charged, with Coulomb focussing increasing the collision rates by
large factors.
Evidently depletion and grain growth can take place on time scales
$\sim 2\times 10^5{\rm yr}$ in ordinary HI clouds, 
with $n_{\rm H}\approx30\,{\rm cm}^{-3}$.

We conclude that the low gas-phase abundances of elements such as Si and Ti
can be understood based on rapid rates of depletion from the gas phase
for the conditions found in ordinary HI clouds -- the ``cold neutral medium''.
The bulk of interstellar grain material is the result of accretion from
the gas in the cold ISM.

\section{What Materials Form?  Why Do Some Elements Not Deplete?}
\subsection{Thermal Desorption}
In typical diffuse clouds, some elements (e.g., Ca, Al, Si, Fe) are heavily
``depleted'' from the gas, 
while others (e.g., S, Cu, Zn) show much less of a tendency to deplete
\citep{Jenkins_2004}.
To try to understand how this selectivity can occur, let's examine the
steps in the depletion process.

Grains with radii $a\ga0.01\mu$m remain very cold in the diffuse ISM, with
temperatures $T\la20$K.  At these temperatures, almost any atom or
ion impinging with
a kinetic energy $\sim0.01$eV is expected
to have a high probability for ``sticking'' to the surface at least
temporarily (where ``temporarily'' means long compared to the
characteristic vibrational period $\sim 10^{-12}$s), with the exception
of the inert gases such as He and Ne.  Normal physisorption via van der Waals
forces will bind species such as S, Cu, and Zn to grain
surfaces with binding energies $B\sim0.05$ -- 0.2eV.  Thermal desorption will
take place with a probability per unit time $\sim\nu_0 \exp(-B/kT)$,
with $\nu_0\approx 10^{12}{\rm s}^{-1}$. 
For $B=0.05\,$eV and a grain
temperature $T=20$K, the lifetime against thermal desorption would
be $\sim1\,$s, whereas if $B=0.10\,$eV, the lifetime becomes $5\times10^5\,$yr.
Therefore if an atom is to remain on the grain surface long enough to matter,
it must have a binding energy $B> 0.1\,$eV.
Unfortunately, the binding energies of the atoms of interest
(C, Mg, Si, Fe) to surfaces of interest (amorphous silicate or carbonaceous
materials) are not known, but it is plausible to imagine that those elements
that do deplete are held to the grains surface by binding energies
$B > 0.1\,$eV.

\subsection{Hydrogenation}
We must now estimate some other rates.
Consider one ``surface site'' on a grain, where some species
X is adsorbed.  The area of this surface site will be 
$A\approx 10^{-15}{\rm cm}^2$.
This surface site is being bombarded by H atoms at a rate
\begin{equation}
n({\rm H})\left(\frac{kT}{2\pi m_{\rm H}}\right)^{1/2}A = 
\frac{1}{30\,{\rm yr}}
\left(\frac{n({\rm H})}{30\,{\rm cm}^{-3}}\right)
\left(\frac{T}{100\,{\rm K}}\right)^{1/2} ~~~.
\end{equation}
The impinging H atoms might form chemical bonds with the adsorbate 
X; whether they
will do so is not known
(for some X there may be an energy barrier in the
``reaction coordinate'' necessary to produce XH, preventing reaction at low
energies) but it does seem possible that incident H might result
in hydrogenation of some physisorbed atoms, such as S.  The resulting molecule
might be ejected from the grain surface, or might remain there.

In H~I clouds, O atoms are the third most abundant species in the gas
(after H and He), and will collide with a surface site at a rate
\begin{equation}
n({\rm O})\left(\frac{kT}{2\pi m_{\rm O}}\right)^{1/2}A
= \frac{1}{3\times10^5{\rm yr}}
\left(\frac{n_{\rm H}}{30\,{\rm cm}^{-3}}\right)
\left(\frac{T}{100\,{\rm K}}\right)^{1/2} ~~~.
\end{equation}

\subsection{Photodesorption or Photolysis by UV Radiation}

Dust grains in diffuse clouds are exposed to diffuse starlight, including
ultraviolet radiation from hot stars.  The starlight intensity is such
that, for example, a C atom in the gas (with ionization energy 11.26 eV) 
will be photoionized at a rate 
$\zeta({\rm C \rightarrow C^+})=3\times10^{-10}\,{\rm s}^{-1}$.
Species on grain surfaces will in general have allowed transitions
to excited states with $E < 13.6~$eV, and these transitions will also
be excited by the starlight background, and we may assume a characteristic
rate for UV excitation
\begin{equation}
\zeta \approx 10^{-10}\,{\rm s}^{-1}=\frac{1}{300\,{\rm yr}}
~~~.
\end{equation}
What happens when the adsorbed species (which might be an atom X,
a hydride XH, an oxide XO, a hydroxide XOH, ...) absorbs a UV photon
and makes a transition to an electronically excited state?  The answer is:
it depends.  Some materials (e.g., stainless steel or quartz) can undergo
heavy ultraviolet irradiation without being affected.  Other materials
(e.g., polyethylene) are chemically altered by UV irradiation.
Some physisorbed or chemisorbed 
species will undergo photoexcitation to an electronic
excited state that may actually be repulsive with respect to the surface,
resulting in ejection in a time $\sim 10^{-13}\,$s after being photoexcited.
Laboratory studies of H$_2$O and mixed ices irradiated by UV radiation
\citep{Westley+Baragiola+Johnson+Baratta_1995b,
       Oberg+Linnartz+Visser+vanDishoeck_2009,
       Oberg+vanDishoeck+Linnartz_2009}
show that the surface erodes at a rate consistent with ejection of
molecules from the surface monolayer with a high yield per photoexcitation
in the surface monolayer.

Comparison of eq.\ (10-12) shows 
that a species X residing on a grain surface will have a good
chance of encountering an impinging 
H atom before anything interesting happens,
but will be very unlikely to encounter any other impinging species
(e.g, C, O, Si) before undergoing photoexcitation.
The adsorbed species will undergo $\sim10^4$ photoexcitations before
a monolayer of material (other than adsorbed H) 
can be deposited on top of it.

Because photoexcitation happens so frequently, 
the only materials that {\it can} form by accretion in the diffuse
ISM are those that are not destroyed or altered by ultraviolet irradiation --
this is what selects for amorphous silicate material and carbonaceous
material as the two condensates.  There may be other materials present, of
course -- e.g., Fe oxides, or even metallic Fe -- but observations of
interstellar dust appear to be consistent with amorphous silicates containing
most of the condensed material that is not carbonaceous.

Ultraviolet irradiation
may also explain how it is possible for the ISM to be
able to grow two distinct grain materials -- carbonaceous material and
amorphous silicate material -- out of a single gas mixture.  One can
imagine the following scenario: Suppose the grain population already
has some amorphous silicate material exposed to the interstellar gas.
When Mg, Si, Fe, and O atoms and ions arrive at the grain surface, they are
able to grow additional amorphous silicate, adding one atom at a time.
It may actually help to have ultraviolet radiation present, as the
electronic excitations may allow chemical bonds to be rearranged to
form amorphous silicate material as
new O, Mg, Si, Fe atoms are added to the surface layer.

On the other hand, what happens when a C atom, for example, lands on the
amorphous silicate surface?  One can imagine that the C atom physisorbed on
the amorphous silicate surface might
undergo photoexcitation to an excited state that is repulsive, ejecting it
from the surface. Or perhaps the C would become hydrogenated or oxidized,
with the resulting CH or CO undergoing photodesorption from the surface.
Such processes could keep the amorphous silicate carbon-free.

Similar processes may occur on exposed carbonaceous surfaces: impinging
C atoms could grow new carbonaceous material, whereas impinging Mg, Si, Fe,
etc.\ could be removed by some combination of reaction with impinging H or O,
and photoexcitation by UV.

In this scenario, we can imagine the carbonaceous material and amorphous
silicate material being in separate grains, or perhaps even as separate
regions of a single grain.  In the latter case, one would expect that
when grain-grain collisions cause grains to shatter, the fragmentation
would preferentially occur along interfaces between the two materials,
leading to ``pure'' fragments.

Perhaps, then, the interstellar grain population actually is dominated by two
distinct grain types: amorphous silicate grains and some form of
carbonaceous material.

\section{Growth of Dust at High $z$: the example of J114816+525150}

Observations of quasars and luminous galaxies at high redshift have
detected large masses of dust in a number of systems
\citep{Wang+Carilli+Wagg+etal_2008}.
The prime example is the galaxy associated with the $z=6.42$ QSO
SDSS J114816+525150 (hereafter J1148+5251), with estimated dust mass 
$M_{\rm dust}$
ranging from $2\times10^8M_{\sun}$
\citep{Dwek+Galliano+Jones_2007a} to
$7\times10^8M_{\sun}$
\citep{Bertoldi+Carilli+Cox+etal_2003}.
Molecular gas appears to account for a large fraction of the 
$\sim5\times10^{10}M_{\sun}$ dynamical mass 
\citep{Walter+Carilli+Bertoldi+etal_2004}.
With $M_{\rm gas}<5\times10^{10}M_{\sun}$, 
and $M_{\rm dust}\ga 2\times10^8M_{\sun}$,
this system has a dust/gas mass ratio
$M_{\rm dust}/M_{\rm gas} > 0.004$ -- comparable to or exceeding the
Milky Way value $\sim$0.005.

The time available for stellar evolution prior to $z=6.42$ is limited:
if star formation began at $z=10$, the oldest stars are only 400 Myr old
at $z=6.42$, and only massive stars will have been able to evolve -- 
low-mass stars have insufficient time to evolve to the asymptotic giant
branch which dominates production of stardust in the Milky Way.
This has led a number of 
authors to propose that supernovae are responsible for the
dust in high-$z$ galaxies
\citep[e.g.][]{Maiolino+Schneider+Oliva+etal_2004,
               Sugerman+Ercolano+Barlow+etal_2006,
               Bianchi+Schneider_2007}.
\citet{Dwek+Galliano+Jones_2007a} discuss the dust in J1148+5251 and
conclude that supernovae would have to produce $\ga1M_{\sun}$ of dust
per supernova to explain the observations, but note that this
is considerably in excess of what has been observed in SN ejecta.

J1148+5251 contains a large mass of molecular gas, detected in
CO\,$J=7\rightarrow6$, $6\rightarrow5$, and $3\rightarrow2$ 
\citep{Bertoldi+Cox+Neri+etal_2003,
       Walter+Carilli+Bertoldi+etal_2004}.
The CO that is
observed, and the H$_2$ that must accompany the CO, is not
supernova-produced: even if those molecules do form in the ejecta, they
are efficiently 
destroyed by the reverse shock when the high-velocity ejecta are decelerated.

Instead, the H$_2$ must be formed by the same mechanism that dominates
H$_2$ formation in the Milky Way: catalysis on grain surfaces.
Given that the massive stars present in these galaxies will destroy
H$_2$ molecules -- primarily through photodissociation -- each H nucleon
in the gas must, on average, have collided with grain surfaces 
many times in the age of
this galaxy.  Metal atoms
and ions move only a few times more slowly than H, and will also
 collide with grain surfaces many times; if they stick,
they will form new grain material.  
As discussed above, this is the process that dominates
grain formation in the Milky Way, and there is no reason not to expect
it to formation of grain material in J1148+5251.

Supernovae are of course required to produce the metals that compose
the grains, and to provide some supernova-condensed ``stardust'' to provide
``seed'' surface area on which to grow more material in the ISM, but the
bulk of the dust mass in high-$z$ galaxies appears likely to be the result
of grain growth competing successfully with grain destruction in the
ISM.

\section{Summary}
%\section{}   %%% Top level section head (remove "%" symbol)
%\subsection{}   %%% Second level section head (remove "%" symbol)
%\subsubsection{}   %%% Lowest level section head (remove "%" symbol)
%\section*{}    %%% Unnumbered top level section head (remove "%" symbol)
%\subsection*{}   %%% Unnumbered second level section head (remove "%" symbol)

The principal points of this paper are as follows:
\begin{enumerate}
\item 
The principal observational constraints on models for the
interstellar grain population are summarized.  The observed wavelength
dependence of extinction (Fig.\ 1) and polarization of starlight, 
together with elemental abundances (total and
observed in the gas) continue to provide the strongest constraints on
what interstellar dust can be.
Additional information is provided by infrared emission features
at 3.3, 6.2, 7.7, 8.6, 11.3, 12.0$\mu$m that are identified as coming
from PAHs following single-photon heating.
\item
A grain model consisting of a population of amorphous silicate spheres
and a population of carbonaceous grains, including PAHs, can reproduce
the observed wavelength-dependent extinction by dust.  This model
also successfully reproduces the observed infrared emission from
interstellar dust (Figs. 2, 3).
\item
If the amorphous silicate grains and carbonaceous grains are given spheroidal
shapes, the observed polarization of starlight can also be reproduced. 
\item
The infrared emission spectra calculated for these 
spheroidal grain models (Fig.\ 4) are
consistent with the observed infrared emission from
high-latitude regions in the Milky Way.
The models predict different
degrees of polarization as a function of wavelength -- observations
with {\it Planck} will test these predictions.
\item
The evolution of interstellar grains in the Milky Way is discussed.
Grain destruction in the ISM is such that $\la$10\%
of the interstellar dust mass consists of ``stardust'' from stellar
sources, including supernovae.
{\it The bulk of interstellar dust has been grown in the ISM}.
\item
The amorphous silicate and carbonaceous materials present in interstellar
grains are the product of grain growth in the presence of
strong ultraviolet radiation, 
that will photoexcite and possibly photodesorb
species from the grain surface.
The ultraviolet radiation may be key to maintaining separate populations
of silicate and carbonaceous grains even though most grain growth occurs
out of an average-composition ISM.
\item
Dust in high-redshift systems such as J1148+5251 must be 
predominantly the result of grain growth in interstellar clouds.
A small amount of dust from supernovae is sufficient to initiate grain
growth in the ISM.
\end{enumerate}

\acknowledgements
I thank the organizers -- particularly Thomas Henning,
Eberhard Gr\"un, and Juergen Steinacker --
for making this conference a reality.
I am grateful to Adam Burrows and Lindsay Keller
for helpful discussions, to the referee for a careful reading,
and to Robert~H.~Lupton for availability of the SM plotting program.
This work was supported in part by NSF grant AST-0406883.

%%% THE BIBLIOGRAPHY
%%%
%%% CONSULT SECTION 3 OF "INSTRUCTIONS FOR AUTHORS" FOR HOW TO USE NATBIB.
%%% AUTHORS ARE ENCOURAGED TO USE EITHER THE "THEBIBLIOGRAPY" ENVIRONMENT
%%% BY UNCOMMENTING (DELETING THE "%" SYMBOL) THE COMMANDS BELOW, OR BY
%%% USING THE BIBTEX ENVIRONMENT. TO FIND OUT WHICH IS APPLICABLE TO YOUR
%%% CONTRIBUTION, CONSULT THE VOLUME EDITORS FOR YOUR PROCEEDINGS.
%%%

%\begin{thebibliography}{}
%\bibitem[]{}
%\bibitem[]{}
%\bibitem[]{}
%\bibitem[]{}
%\bibitem[]{}
%\bibitem[]{}
%\bibitem[]{}
%\bibitem[]{}
%\bibitem[]{}
%\bibitem[]{}
%\bibitem[]{}
%\bibitem[]{}
%\end{thebibliography}

\bibliography{btdrefs}

\end{document}